\documentclass[11pt]{article}
\usepackage{epsf,graphics}

\def\beq{\begin{equation}}
\def\eeq{\end{equation}}
\date{\today}

\begin{document}
\begin{center}
{\large\bf The $\epsilon$ expansion and Universality in three dimensions}\\[.3in]
  {\bf   Nicolas Sourlas} \\
  Laboratoire de Physique Th\'eorique de l'ENS, \'Ecole Normale Sup\'erieure, 
  PSL Research University, Sorbonne Universit\'es, UPMC Univ. Paris 06, CNRS, 75005 Paris, 
  France.
 
\ 

\end{center}
\vskip .15in
\centerline{\bf ABSTRACT}

It has been observed that the clasification into universality classes 
of critical behaviour, as established by perturbative renormalization group 
in the viscinity of four or six dimensions of space by the epsilon expansion,  
remains valid down to three dimensions in all known cases, 
 even when purturbative renormalisation group 
fails in three dimensions. In this paper we argue that this classification 
into universality classes remains true in lower dimensions of space, 
even when purturbative renormalisation group fails,
because of the well known phenomenon of eigenvalue repulsion.

\begin{quotation}

\vskip 0.5cm
\noindent
PACS numbers: 05.10Cc,05.70.Jk

\end{quotation}

One of the big triumphs of the renormalization group(RG) is the explanation 
of universality near a second order phase transition\cite{WK,W}. 
Very different physical systems like 
uniaxial magnets, binary mixtures, liquid-vapor transitions, 
 share the same values of 
critical exponents, critical amplitude ratios etc. We say they belong to 
the same universality class. 
Few important parameters like the dimensionality of space or the symmetry of 
the system, determine the universality class. Details of the interaction are 
irrelevant. 

Near a second order phase transition, the hamiltonian of the 
system flows under renormalization group transformations, toward a fixed point. 
In the viscinity of this 
fixed point one can use Wilson's operator product expansion\cite{ope}  
and analyze the dimensions of the operators appearing in the Hamiltonian.
The operators are classified according to their scaling dimension 
into relevant, marginal and irrelevant.

Physical systems whose hamiltonians differ by irrelevant operators 
belong to the same universality class.

In general it is not easy to find the RG fixed point.
But above a certain dimension of space $D_u$, called the upper critical 
dimension, the renormalization group flows to the gaussian fixed point and 
mean field theory is valid. 
$D_u=4$ for systems with $Z_2$ symmetry like Ising ferromagnets 
and $D_u=6$ for systems where cubic operators are allowed, like Potts models. 

Near the upper critical dimension the computation of the dimensions of operators can 
easily be performed in the framework of the $\epsilon= D_u-D$ 
expansion\cite{FW}. $D$ is the dimension of space. 
This allows for the classification into universality classes 
of physical systems for infinitesimal $\epsilon $.

What is quit remarquable is that this classification remains valid 
at $D=3$, i.e. for $\epsilon = 1 $ or $\epsilon = 2 $. 
Indeed there are not known cases of the classification into 
universality classes, based on the epsilon expansion,  
not being valid at D=3.

This fact is even more remarquable in the case of disordered systems. 
The $\epsilon $ expansion is based on the perturbative renormalization group.
The straightforward application of
field theoretic methods and the renormalization group (RG) is not
possible because the disorder breaks the translation symmetry of
the Hamiltonian. The standard procedure is then to average over
disorder using the replica method~\cite{EA}. One starts
with $n$ noninteracting copies of the system (replicas) and
averages over the disorder distribution. This produces an
effective Hamiltonian with $n$ interacting fields which is
translation invariant and enables the use of the RG. In the end,
the $ n \to 0 $ limit has to be taken.
The replica method is mathematically unorthodox. Its combination
with the perturbative renormalization group (PRG) has been shown
to produce incorrect results in 3D systems. An example is
provided by the random-field Ising model (RFIM) where the
combination of the replica method with the PRG predicts
dimensional reduction~\cite{Y,PS},
which does not hold neither in three~\cite{BI,BK} nor in
four dimensions~\cite{F4D}. Mean field and the
replica method are believed to be correct at infinite dimensions.

Despite this failure of PRG it has beeng shown that, contrary to previous believes, 
universality holds for the RFIM in three dimensions. 
PRG predicts that different RFIM models, where the 
random fields are drawn from different probability distributions of the 
random fields, belong to the same 
universality classes. Also, more surprisenly, diluted antiferromagnets in a field 
are predicted to belong to the same universality class. These two predictions 
have been recently shown numerically to be true in three and four dimensions~\cite{FMM,PIS}, 
despite the failure of PRG

Why even when perturbative renormalization group fails, its predictions about 
universality classes still remain true at low dimensions?

In this note we will argue that if perturbative renormalization group is valid 
in the viscinity of the upper critical dimension, i.e. if the epsilon expansion is 
valid for infinitesimaly small 
$\epsilon$, physical systems which according to 
the  $\epsilon$ expansion belong to the same universality class, continue to 
belong to the same universality class also in lower dimensions, 
even when PRG is not valid anymore 
at these lower dimensions. 

The reason has to do with the scaling dimensions of the operators 
which appear in Wilson's operator product expansion and 
is the following. The renormalization group fixed point 
and the scaling dimension $d_O $ of the operator $O $'s 
are a function of the dimension of space $D$. The $d_O(D) $'s  change when the 
dimension of space is changed. A necessary and 
sufficiant condition 
for non changing universality classes as the dimension of space $D$ varies is 
for the 
scaling dimensions $d_O(D) $ of the leading operators not to cross when the 
dimension of space is lowered  
from $D=D_u$ down to $D=3$ as this is illustrated in the figure. 
If this is the case, the classification of the operators into relevant, marginal 
and irrelevant remains unchanged when the dimension of space is lowered.
The essential observation is that 
the  scaling dimensions $d_O(D) $'s of the operators 
are eigenvalues of the scaling transformations, 
i.e. of the group of dilatations of space. 
It is well known from the early days of quantum mechanics that in 
the generic case the eigenvalues of operators, i.e. the eigenvalues of the 
matrices in the matrix representation 
of the operators, in the case of quantum mechanics the Hamiltonian of the system, 
do not cross if one changes a single parameter\cite{LL}. 
This is easily verified for a two by two hermitian matrix, 
which depends on a parameter $s$. 
The condition for the two eigenvalues of the matrix to be equal is 
 that a sum of two perfect squares $ a(s)^2 + b(s)^2  $ is zero,
 i.e. $a(s)=0 $ and $b(s)=0 $. In the generic case it is not possible to 
 satisfy simultaneously these two equations by fixing a single parameter $s$.
The general mathematical argument is due to von Neumann and Wigner\cite{WVN}.
This phenomenon of eigenvalue repulsion is the reason for which universality 
classes do not change when the dimension of space is lowered. It does not 
depend on the validity of perturbative renormalization group at lower dimensions.
There is a regularity asumption in the previous argument, i.e. that one can follow 
continuously the change of the scaling dimensions of the operators when the 
dimension of space is lowered.

The previous argument can be inverted. If it is found by other means, 
like experiments or numerical simulations, that 
the epsilon expansion classification of 
universality classes is valid at three dimensions, it means that perturbative 
renormalization group is valid near the upper critical dimension $D_U$, 
i.e. the epsilon expansion is valid. This is a non trivial information in  
the case of disordered systems.

It has been shown recently\cite{FMMPS} that dimensional reduction 
as predicted by the perturbative renormalization group for the 
random field Ising model\cite{Y,PS}, is indeed valid at five dimensions. 
The breaking of perturbative renormalization group is a low dimensional phenomenon 
and does not affect universality.
\newpage
{\epsfxsize=12cm
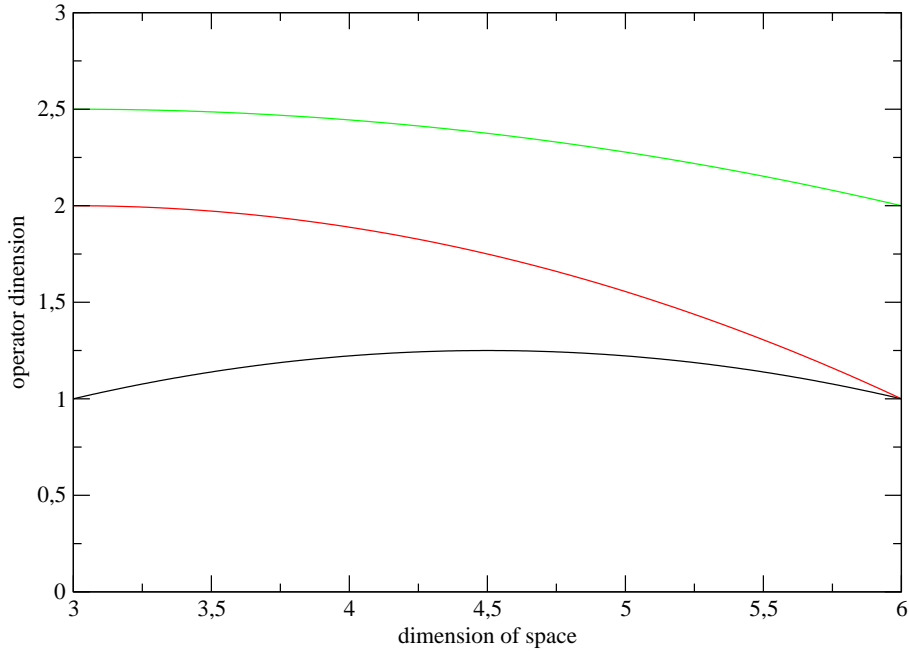
\epsffile{fig0.eps}}

$$ \ $$
Useful and stimulating discussions with Ed Witten are gratefully aknowledged.

\end{document}